\documentclass[12pt]{iopart}

\usepackage{graphicx}
\usepackage{dcolumn}
\usepackage{bm}
\usepackage{hyperref}
\usepackage{amssymb}
\usepackage{amsfonts}

\newcommand{\meV}{{\rm meV}}
\newcommand{\eV}{{\rm eV}}

\newcommand{\GeV}{{\rm GeV}}
\newcommand{\TeV}{{\rm TeV}}

\newcommand{\s}{{\rm s}}

\newcommand{\Mpl}{M_{\rm Pl}}

\begin{document}


\title{GZK photon constraints on Planck scale Lorentz violation in QED}

\author{Luca Maccione, Stefano Liberati} 

\address{SISSA, Via Beirut, 2-4, I-34014, Trieste, Italy} 
\address{INFN, Sezione di Trieste, Via Valerio, 2, I-34127, Trieste, Italy}

\eads{\mailto{maccione@sissa.it}, \mailto{liberati@sissa.it}}


\begin{abstract}
We show how the argument exploited by Galaverni \& Sigl in Phys.~Rev.~Lett., 100, 021102 (2008) (see also arXiv:0708.1737) \cite{Galaverni:2007tq} to constrain Lorentz invariance violation (LV) using Ultra-High-Energy photon non observation by the AUGER experiment, can be extended to QED with Planck-suppressed LV (at order $O(E/M)$ and $O(E^2/M^2)$).
While the original constraints given by Galaverni \& Sigl \cite{Galaverni:2007tq} happen to be weakened, we show that, when used together with other EFT reactions and the expected detection of photons at $E > 10^{19}$ eV by AUGER, this method has the potentiality not only to basically rule out order $O(E/M)$ corrections but also to strongly constrain, for the first time, the CPT-even $O(E^2/M^2)$ LV QED.
\end{abstract}


\section{Introduction}
\label{sec:intro}

The interest in possible high energy violations of local Lorentz Invariance (LI) has grown in recent years, encouraged by the flourishing of observational tests and the availability of experimental data. Theoretically, hints of Lorentz Violation (LV) arose from various approaches to Quantum Gravity (QG) \cite{KS89,AmelinoCamelia:1997gz,LoopQG,Carroll:2001ws, Lukierski:1993wx, AmelinoCamelia:1999pm,Burgess:2002tb,Analogues}.


While it may seem hopeless to search directly for effects suppressed by the Planck scale, it is possible that even tiny corrections can be magnified to measurable ones when dealing with high energies, long distances of propagation or peculiar reactions (see e.g.~\cite{Mattingly:2005re, AmelinoCamelia:2002dx}). However, in order to do so, a complete theoretical framework, within which it is possible to calculate reaction rates and to describe the particle dynamics, is needed.

This is the case of Effective Field Theory (EFT) with LV operators. We shall deal here with modified QED via non-renormalizable, Planck suppressed LV operators (the analogue theory with renormalizable operators being already severely constrained \cite{Mattingly:2005re}). It has been shown \cite{Myers:2003fd,Mattingly:2008pw} that the addition of the two lowest order non-renormalizable LV operators (mass dimension 5 and 6 respectively) to the effective Lagrangian of QED leads to the following high-energy modified dispersion relations (MDR) (since the LV correction is proportional to $p^{n}$ we call them $n-$MDR with $n=3$ for dimension 5 operators and $n=4$ for dimension 6 ones)
\begin{eqnarray}
\omega_{\pm}^2 &=& k^2 + \xi^{(n)}_{\pm} k^n/M^{n-2}
\label{eq:disp_rel_phot}\\
 E_\pm^2 &=& p^2 + m_{e}^2 + \eta^{(n)}_\pm p^n/M^{n-2}\;,
\label{eq:disp_rel_ferm}
\end{eqnarray}
where (\ref{eq:disp_rel_phot}) refers to photons
while (\ref{eq:disp_rel_ferm}) refers to fermions~\footnote{
Actually, $4-$MDR contains an extra term, proportional to $p^{2}$, due to CPT even dimension 5 operators~\cite{Mattingly:2008pw}.  However this term is suppressed by $m_{e}/\Mpl$.}. 
(In the following we will assume $M$ to be comparable to the Planck mass $M_{\rm Pl} \simeq 1.22\times 10^{19}~\GeV$.) 
The constants $\xi^{(n)}_{\pm}$ and $\eta^{(n)}_\pm$ indicate the strength of the LV and take values on the whole real axis. In (\ref{eq:disp_rel_phot}) the $+$ and $-$ signs denote right and left circular polarization, while in (\ref{eq:disp_rel_ferm}) they indicate opposite helicity fermion states. 

A crucial difference between the $n=3$ and $n=4$ cases is the fact that the former is characterized by LV terms which break CPT invariance, while the relevant ones for the $4-$MDR are CPT even \cite{Mattingly:2008pw}. This difference implies that for the $3-$MDR there is an effective breaking of the symmetry between the two helicity states of the photon. Indeed one finds that $\xi^{(3)}_{+} = -\xi^{(3)}_{-} \equiv \xi^{(3)}$, while $\xi^{(4)}_{+} = \xi^{(4)}_{-} \equiv \xi^{(4)}$~\cite{Mattingly:2008pw}.  
On the other hand, it can be shown that the coefficients of electrons and positrons are related as $\eta_{\pm}^{e^{-}} = (-)^{n}\eta_{\mp}^{e^{+}}$, exploiting the argument given in \cite{Jacobson:2005bg, Mattingly:2008pw}.

Since suitable powers of the suppressing scale $\Mpl$ have been already factored out in  eq.(\ref{eq:disp_rel_phot}, \ref{eq:disp_rel_ferm}), natural values of the LV, dimensionless coefficients in (\ref{eq:disp_rel_phot}, \ref{eq:disp_rel_ferm}) are expected to be $O(1)$.

 From a purely logical point of view, it could seem unreasonable to study $O(E^2/M^2)$ LV corrections (those leading to $4-$MDR), as these will be always subdominant with respect to those at $O(E/M)$ (those leading to $3-$MDR). However, the reasons for this interest are both empirical and theoretical. 

On the observational side, the LV parameters $\eta^{(3)}_{\pm}$ are presently constrained to be less than $O(10^{-5})$ at 95\% confidence level (CL) by a detailed analysis of the synchrotron component of the Crab Nebula broadband spectrum \cite{Maccione:2007yc}, while the constraint $|\xi^{(3)}| \lesssim 10^{-7}$ is obtained by considering the absence of vacuum birefringence effects in the propagation of optical/UV polarized light from Gamma-Ray Bursts~\cite{Fan:2007zb}.

On the theoretical side, a reasonably good motivation for focusing on $O(E^2/M^2)$ LV corrections is related to the so called ``naturalness problem" \cite{Jacobson:2005bg}. In fact, it is generic that, even starting with an EFT with only mass dimension 5 or 6 LV operators for free particles, radiative corrections due to particle interactions will generate lower dimension LV terms without introducing any further suppression~\cite{Collins:2004bp}. Hence extra LV terms in $p$ and $p^2$ will be generically dominant on higher order LV ones and lead to extremely stringent constraints on the dimensionless LV coefficients $\xi^{(n)},\eta^{(n)}$. 

However, it has been shown~\cite{Bolokhov:2005cj} that if the theory includes SuperSymmetry (SUSY), then dimension 3 and 4, renormalizable, LV operators are forbidden. As a consequence, renormalization group equations for Supersymmetric QED with dimension 5 LV operators \`a la Myers \& Pospelov were shown not to generate lower dimensional operators, if SUSY is unbroken. 
%

SUSY soft-breaking does lead to LV terms in $p$ and $p^2$, however characterized only by a suppression of order $m_{s}^{2}/\Mpl$ ($n=3$) or $(m_{s}/\Mpl)^{2}$ ($n=4$), where $m_{s}\simeq 1~\TeV$ is the scale of SUSY soft breaking \cite{Bolokhov:2005cj}.  Nonetheless, given the present constraints, dimension 5 LV operators would induce dimension 3 ones which are already tremendously constrained. Hence, in the $n=3$ case one would have to require unnaturally small LV coefficients in order to have a viable model. On the contrary, if $n = 4$ then the induced dimension 4 terms are suppressed enough to be compatible with current constraints without requiring $\xi^{(4)},\eta^{(4)}$ much less than one, provided $m_{s} < 100~\TeV$.
Therefore, missing an alternative ``custodial symmetry" for LV with respect to SUSY, QED dimension 5 LV operators seem problematic, while dimension 6 LV, CPT even, ones are favored. 

At present, a clear general argument, as to why dimension 5 LV operators should not appear, is missing. However, if we assume, together with SUSY symmetry with $m_{s} < 100~\TeV$, also CPT invariance for the Planck scale theory, then not only dimension 3 and 4, but also dimension 5, CPT odd, LV operators would be forbidden and only CPT even, dimension 6 ones would appear in the effective Lagrangian \cite{Mattingly:2008pw}~\footnote{It is however important to stress that the SUSY LV operators considered in \cite{Bolokhov:2005cj} do not lead to dispersion relations of the form presented here.}. 

Given this state of affairs, we shall provide here constraints on both dimension 5 and dimension 6 operators separately.
Of course, in order to do so for dimension 6 operators one would need to explore much higher energies that those associated to the above cited constraints on $n=3$ dispersion relations. This is why we shall focus here on the physics of ultra-high-energy cosmic rays (UHECR).

It has been recently pointed out \cite{Galaverni:2007tq} that if Lorentz symmetry was violated, then the absorption of ultra-high-energy (UHE) photons ($E > 10^{19}~\eV$) on the Cosmic Microwave Background (CMB) and the Universal Radio Background (URB) could be forbidden, thus leading to large photon fluxes reaching Earth. This would violate limits put by current experiments \cite{Aglietta:2007yx,Rubtsov:2006tt} on the photon fraction in UHECR. Hence, very strong constraints $|\xi^{(3)}| \lesssim 10^{-15}$ and $\xi^{(4)} \gtrsim -10^{-7}$ were claimed \cite{Galaverni:2007tq}. An underlying assumption in \cite{Galaverni:2007tq} is that $\eta^{(n)} \simeq 0$ in order to prevent competing reactions with respect to photon absorption. This is an important limitation, as we will show below.

The main result of this work is to extend the idea given in \cite{Galaverni:2007tq} to the full LV QED framework described by equations (\ref{eq:disp_rel_phot}) and (\ref{eq:disp_rel_ferm}). While the original constraints \cite{Galaverni:2007tq} are weakened, we shall see that, when used together with other EFT reactions, this method has the potentiality not only to basically rule out the $n=3$ case but also to strongly constrain, for the first time, the CPT-even (hence possibly theoretically favored) $n=4$ LV QED.

This paper is structured as follows: in section \ref{sec:UHEphotons} we briefly describe the origin of the Ultra-High-Energy photons and explain the argument by \cite{Galaverni:2007tq}. In section \ref{sec:LVreactions} we describe the generalization of the previous argument in Effective Field Theory LV. In section \ref{sec:constraints} we discuss our results and draw conclusions.


\section{UHE photons and LV}
\label{sec:UHEphotons}

UHE photons originate in the interactions of UHECRs with the CMB, leading to the production of neutral pions which subsequently decay into photon pairs. Pion production occurs only if the interacting UHECR energy is above $E_{\rm th} \simeq 5\times 10^{19}~(\omega_{b}/1.3~\meV)^{-1}~\eV$ ($\omega_{b}$ is the target photon energy). Hence, it has long been thought to be responsible for a cut off in the UHECR spectrum, the Greisen-Zatsepin-Kuzmin (GZK) cut-off \cite{gzk}. 

Experimentally, the presence of a suppression of the UHECR flux has been confirmed only recently with the observations by the HiReS detector \cite{Abbasi:2007sv} and the Pierre Auger Observatory (PAO) \cite{Roth:2007in}. Although the cut off could be also due to the finite acceleration power of the UHECR sources, the fact that it occurs at the expected energy favors the GZK explanation. The results shown in \cite{Cronin:2007zz} further strengthen this hypothesis.

The PAO and the Yakutsk and AGASA experiments also imposed limits on the presence of photons in the UHECR spectrum. In particular, the photon fraction is less than 2.0\%, 5.1\%, 31\% and 36\% (95\% C.L)~at $E = 10$, 20, 40, 100 EeV  respectively \cite{Aglietta:2007yx,Rubtsov:2006tt}. Although its theoretical computation is quite uncertain and depends on many unknowns related to source and propagation effects \cite{Aglietta:2007yx}, it is established that photons are mainly attenuated by pair production onto CMB and URB.

However, pair production is strongly affected by LV. In particular, the (lower) threshold energy can be slightly shifted and in general an upper threshold (a finite energy above which pair production is no more allowed by energy-momentum conservation) can be introduced \cite{Jacobson:2002hd}. Therefore, if the upper threshold energy happens to be lower than $10^{19}~\eV$, then UHE photons are no more attenuated by the CMB and can reach the Earth constituting a significant fraction of the total UHECR flux,  thereby violating present experimental limits\footnote{This conclusion could be evaded if the GZK process was not effective. However, the large mass difference between pions and electrons implies that, at comparable energies and LV coefficients, the GZK reaction must be much less affected than pair production. Moreover, it can be shown \cite{Galaverni:2007tq} that LV does not affect the kinematics of $\pi^{0}$ decay.} \cite{Galaverni:2007tq}.

However, this argument is not stringent enough to cast constraints on LV in EFT, because in this framework two competitive processes, forbidden in LI physics, are allowed and can effectively dump the photon flux: photon decay in vacuum and photon splitting ($\gamma\rightarrow N\gamma$). In \cite{Galaverni:2007tq} the special case $\eta^{(n)}\sim 0$ and $\xi^{(n)} < 0$ was considered, in order to prevent these extra processes. In the following we will study the full parameter space.


\section{LV reactions}
\label{sec:LVreactions}

In order to perform a consistent analysis in LV EFT, we have to consider three processes related to photon attenuation: pair production, $\gamma$-decay and photon splitting. 
\begin{description}
\item[Pair production]
This well known process occurs whenever the center-of-mass energy of the $\gamma\gamma$ system is sufficient to produce a $e^{+}/e^{-}$ pair, i.e.~it is larger than $2m_{e}$. 
This condition corresponds to $k \geq k_{\rm th} = m^{2}_{e}/\omega_{b}$. 

In the following, we exploit the above mentioned relation $\eta_{\pm}^{e^{-}} = (-)^{n}\eta_{\mp}^{e^{+}}$, as on average the initial state is unpolarized (see however \cite{Galaverni:2008yj} for further discussion on this point). This is justified as $\pi^{0}$ decay produces photons with opposite polarization and the CMB is unpolarized on average. Moreover, the interaction at threshold must occur in $S$-wave, because the particles' momenta need to be aligned \cite{Jacobson:2002hd}. Nevertheless, it is possible that, if the $S$-wave channel is forbidden, higher partial mode interactions occur in off-threshold configuration. In this case, however, the reaction rate is suppressed by partial mode suppression and because only suitably polarized initial states can contribute to it.

Within this framework, and exploiting energy-momentum conservation, the kinematics equation governing pair production is the following \cite{Jacobson:2005bg}
\begin{equation}
\frac{m^{2}}{k^{n}y \left(1-y\right)} =  
  \frac{4\omega_{b}}{k^{n-1}} + \tilde{\xi} - \tilde{\eta} \left( y^{n-1}+(-)^{n}\left(1-y\right)^{n-1}\right)
   \label{eq:ggscat}
\end{equation}
where $\tilde{\xi}\equiv\xi^{(n)}/M^{n-2}$ and $\tilde{\eta}\equiv\eta^{(n)}/M^{n-2}$ are respectively the photon's and electron's LV coefficients divided by powers of $M$, $0 < y < 1$ is the fraction of momentum carried by either the electron or the positron with respect to the momentum $k$ of the incoming high-energy photon and $\omega_{b}$ is the energy of the target photon (we will assume in the following $\omega_{b} = \omega_{\rm CMB}\simeq 6\times 10^{-4}~\eV$ and will not consider pair production onto the URB, as our main conclusions can be drawn using just CMB).  
Note that the symmetry of (\ref{eq:ggscat}) under the exchange $e^{+}\leftrightarrow e^{-}$ is manifested in its symmetry under $y \leftrightarrow 1-y$ and $\tilde{\eta}\leftrightarrow (-)^{n}\tilde{\eta}$.

Pair production can be severely affected by LV. In particular, it has been shown \cite{Jacobson:2002hd} that, rather surprisingly, not only the threshold energy $k_{\rm th}$ is modified, but also an upper threshold is introduced. Physically, this means that at sufficiently high momentum the photon does not carry enough energy to create a pair and simultaneously conserve energy and momentum. However, an upper threshold can only be found in regions of the parameter space in which the $\gamma$-decay is forbidden, because if a single photon is able to create a pair, then {\em a fortiori} two interacting photons will do \cite{Jacobson:2002hd}. 

The structure of the lower and upper thresholds for $n = 3,4$ has been studied in \cite{Jacobson:2002hd} if $\eta_{+} = \eta_{-}$. The same kind of analysis can be extended to the full EFT case. However, being the computation rather cumbersome, we shall evaluate it numerically.
The structure of the constraint is different depending on $n$. If $n = 3$, since $\xi_{+}^{(3)} = -\xi_{-}^{(3)}$ and the constraint is imposed on both left and right polarized photons, it is symmetric with respect to $\xi \leftrightarrow -\xi$.
If $n = 4$ such a symmetric structure is lost. 
\item[$\gamma$-decay] While forbidden in LI theory, this reaction is allowed in a LV framework if the photon energy is above a certain threshold. The latter can be easily derived by solving the relative energy-momentum conservation equation which can be readily inferred from (\ref{eq:ggscat}) by noticing that it corresponds to the limit $\omega_{b}\rightarrow 0$.
\item[Photon splitting] This is forbidden for $\xi^{(n)}<0$ while it is always allowed if $\xi^{(n)} > 0$ \cite{Jacobson:2002hd}. When allowed, the relevance of this process is simply related to its rate. The most relevant cases are $\gamma\rightarrow \gamma\gamma$ and $\gamma\rightarrow3\gamma$, because processes with more photons in the final state are suppressed by more powers of the fine structure constant. 

The $\gamma\rightarrow\gamma\gamma$ process is forbidden in QED because of  kinematics and C-parity conservation. In LV EFT neither condition holds. However, we can argue that this process is suppressed by an additional power of the Planck mass, with respect to $\gamma\rightarrow 3 \gamma$. In fact, in LI QED the matrix element is zero due to the exact cancellation of fermionic and anti-fermionic loops. In LV EFT this cancellation is not exact and the matrix element is expected to be proportional to at least $(\xi E/\Mpl)^{p}$, $p>0$, as it is induced by LV and must vanish in the limit $\Mpl \rightarrow \infty$. 

Therefore, we have to deal only with $\gamma\rightarrow 3 \gamma$. This process has been studied in \cite{Jacobson:2002hd,Gelmini:2005gy}. In particular, in \cite{Gelmini:2005gy} it was found that, if the ``effective photon mass'' $m_{\gamma}^{2} \equiv \xi E_{\gamma}^{n}/\Mpl^{n-2} \ll m_{e}^{2}$, then the splitting lifetime of a photon is approximately $\tau^{n=3}\simeq 0.025\,\xi^{-5} f^{-1}\left(50~\TeV/E_{\gamma}\right)^{14}~\s$, where $f$ is a phase space factor of order 1. 
This rate was rather higher than the one obtained via dimensional analysis in \cite{Jacobson:2002hd} because, due to integration of loop factors, additional dimensionless contributions proportional to $m_{e}^{8}$ enhance the splitting rate at low energy. 

This analysis, however, does not apply to our case, because for photons around $10^{19}$ eV $m_{\gamma}^{2} \gg m_{e}^{2}$ if $\xi^{(3)} > 10^{-17}$ and $\xi^{(4)} > 10^{-8}$. Hence the above mentioned loop contributions are at most logarithmic, as the momentum circulating in the fermionic loop is much larger than $m_{e}$. Moreover, in this regime the splitting rate depends only on $m_{\gamma}$, the only energy scale present in the problem.

We then expect the analysis proposed in \cite{Jacobson:2002hd} to be correct and we infer that the splitting time scale at $E_{\gamma} \simeq 10^{19}~\eV$ is larger than the propagation one (100 Myr for GZK photons) if $\xi^{(3)} < 0.08$, while if $n=4$ it is well above 100 Myr even for $\xi^{(4)} \sim O(1)$.
\end{description}


\section{Results and conclusions}
\label{sec:constraints}

We have demonstrated so far that only $\gamma$-decay and pair production are relevant to our analysis. By considering these processes three kinds of constraints are possible. 

\begin{figure}[tbp]
 \includegraphics[scale=0.4]{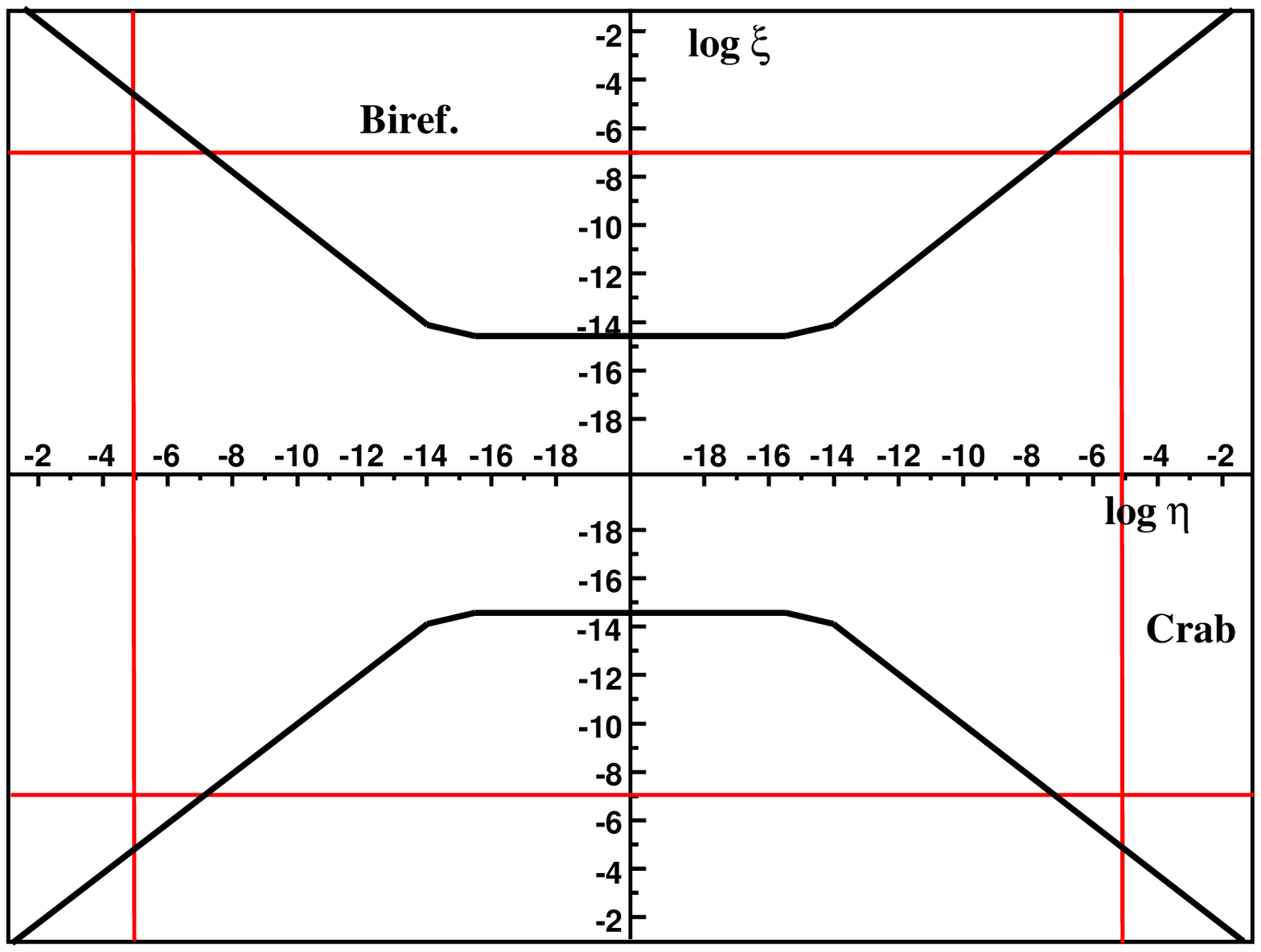}
 \includegraphics[scale=0.4]{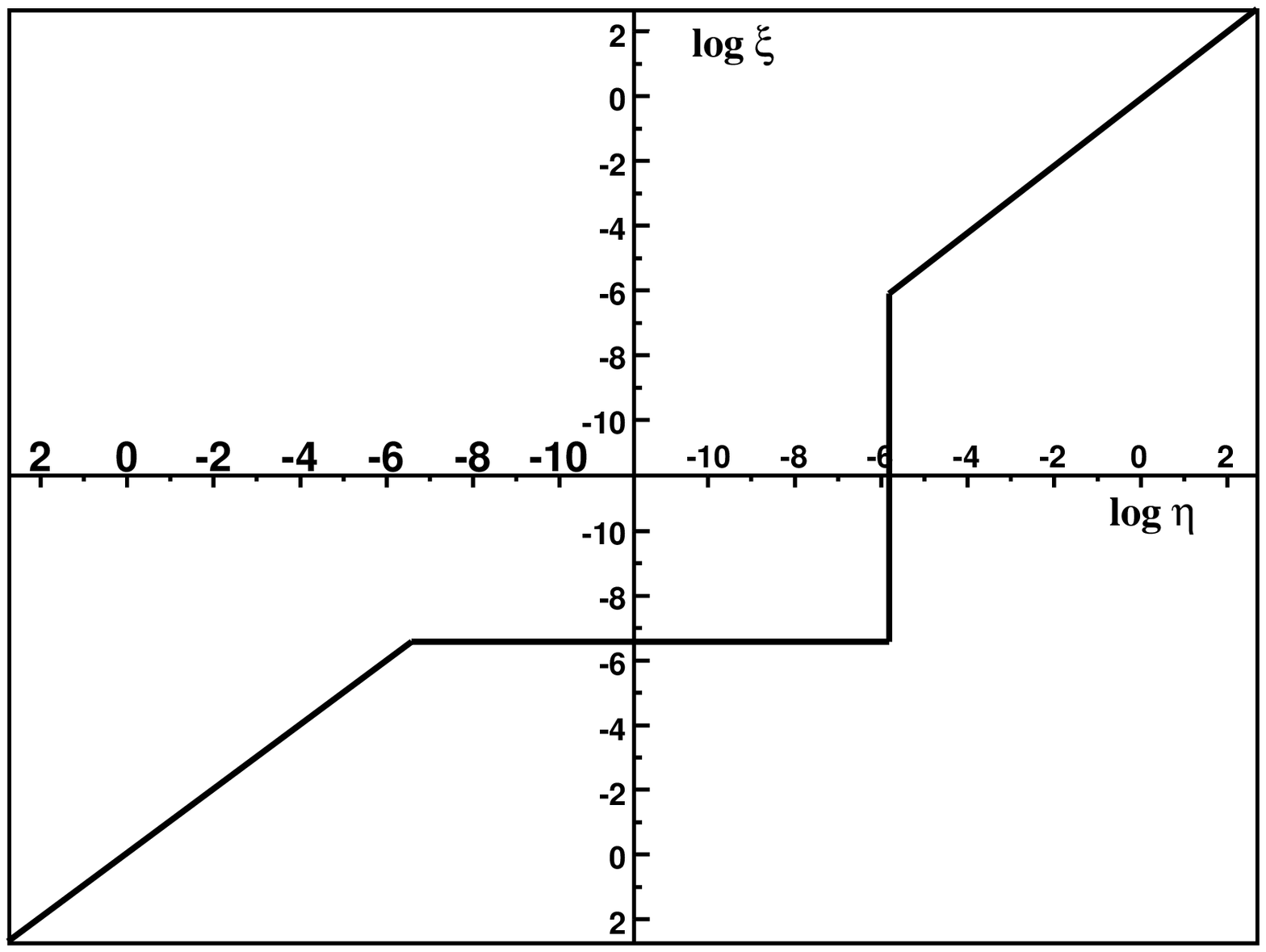}
  \caption{Left panel: $n=3$ LV. Right panel: $n=4$ LV. Constraints from the absence of pair production upper threshold. The best constraints to date are shown in red, if they exist. The allowed region includes the origin and corresponds to the intersection of the regions bounded by the red and black lines.}
 \label{fig:pair3}
\end{figure}

On the one hand, since at present we have stringent upper bounds on photon fluxes up to $10^{20}$ eV \cite{Rubtsov:2006tt}, then any upper threshold energy introduced by LV in pair production must be larger than this figure. This leads to the constraint represented by the black thick solid lines in Fig.~\ref{fig:pair3}, where the allowed region is obviously the one including the origin and in the case $n=3$ is the intersection of the upper threshold allowed region with the ones allowed by already existing constraints (red lines). We confirm the claim by \cite{Galaverni:2007tq} that, for $\eta^{(3,4)} = 0$, $|\xi^{(3)}| \lesssim 10^{-15}$ and $\xi^{(4)}\gtrsim - 10^{-7}$. 
However, Fig.~\ref{fig:pair3} shows that this is a rather special (and favorable) case. 

\begin{figure}[tbp]
  \includegraphics[scale=0.4]{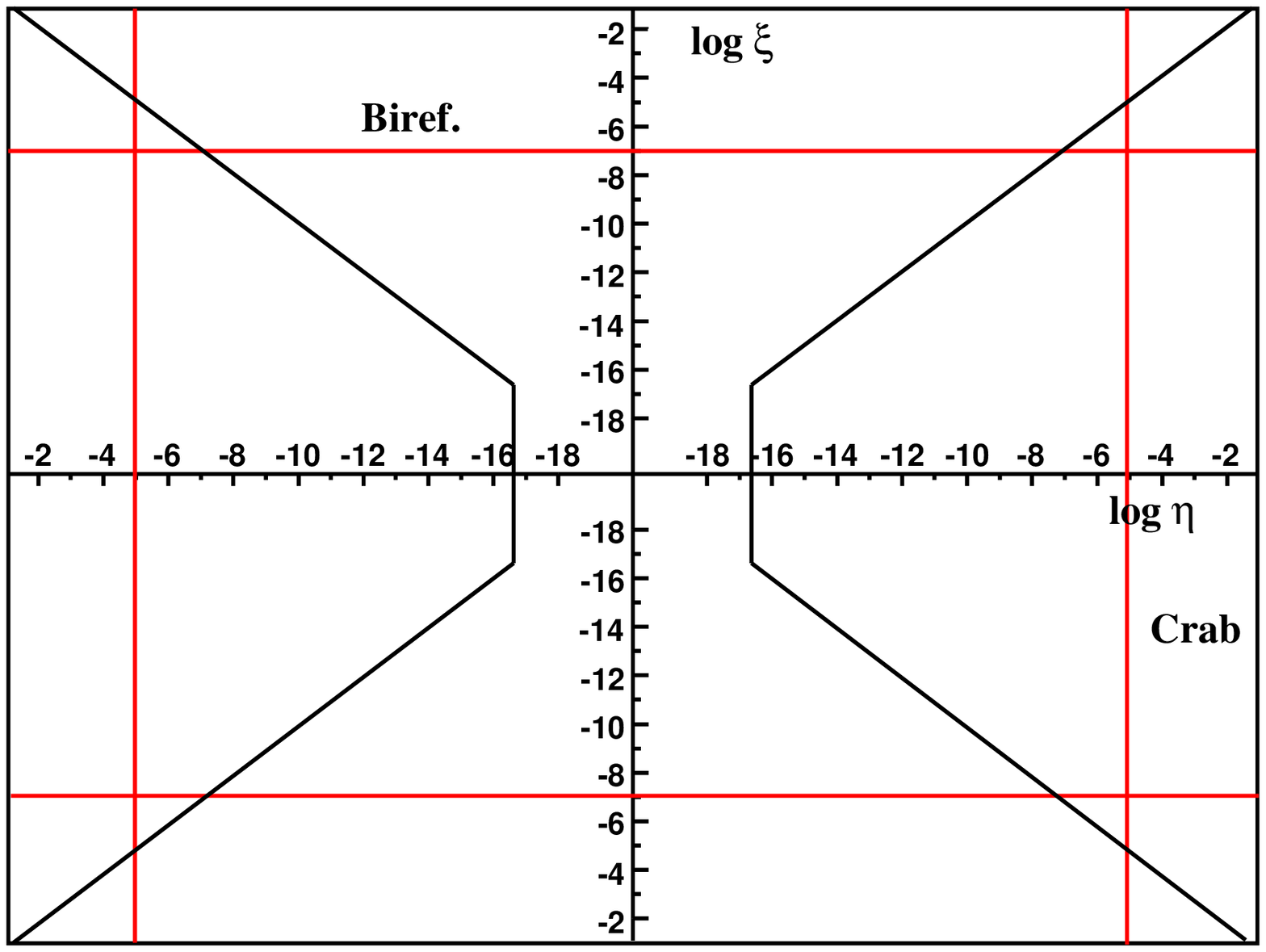}
  \includegraphics[scale=0.4]{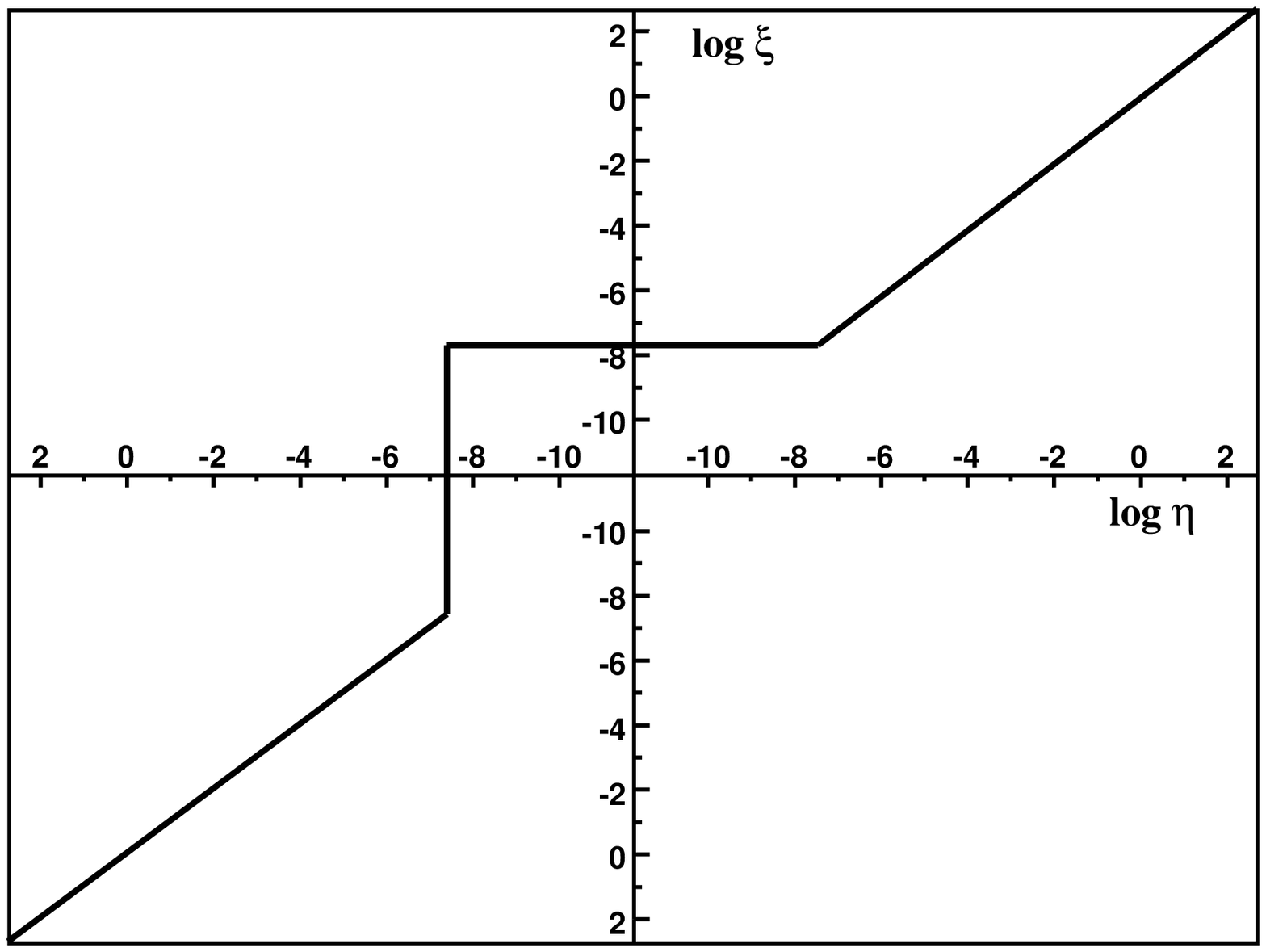}
  \caption{Left panel: $n=3$ LV. Right panel: $n=4$ LV. $\gamma$-decay threshold structure. The best constraints to date are shown in red, if they exist.}
  \label{fig:gdec}
\end{figure}

On the other hand, if some photons were detected above $10^{19}~\eV$, then it could be deduced that the threshold energy for $\gamma$-decay is larger than this energy (once allowed, photon decay is basically instantaneous \cite{Jacobson:2005bg}). Indeed, the PAO will reach the required sensitivity to probe such theoretically expected fluxes \cite{Gelmini:2005wu} within the next few years \cite{Aglietta:2007yx}.
In this case the allowed region for $n=3$ would be the ``clepsydra'' delimited by the black solid line in Fig.~\ref{fig:gdec} and by the two horizontal red lines corresponding to the birefringence constraint. 
As it can be inferred from Fig.~\ref{fig:gdec}, for $n=3$ the combination of the above mentioned constraints would cast the very strong bound $|\eta^{(3)}|< 10^{-7}$. 
Conversely, no significant limit would be placed for $n=4$.
%

\begin{figure}[!thbp]
 \includegraphics[scale=0.4]{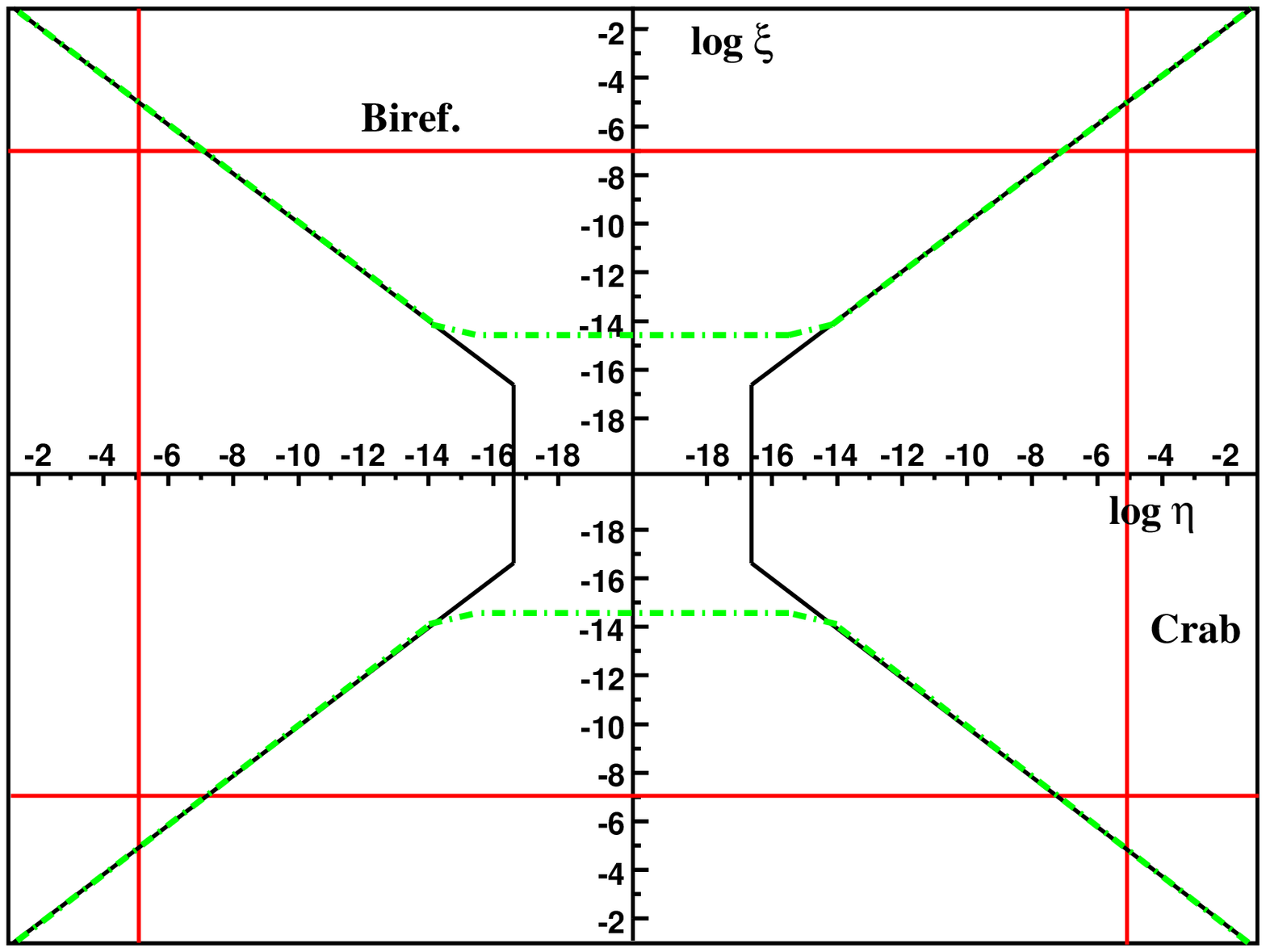}
 \includegraphics[scale=0.4]{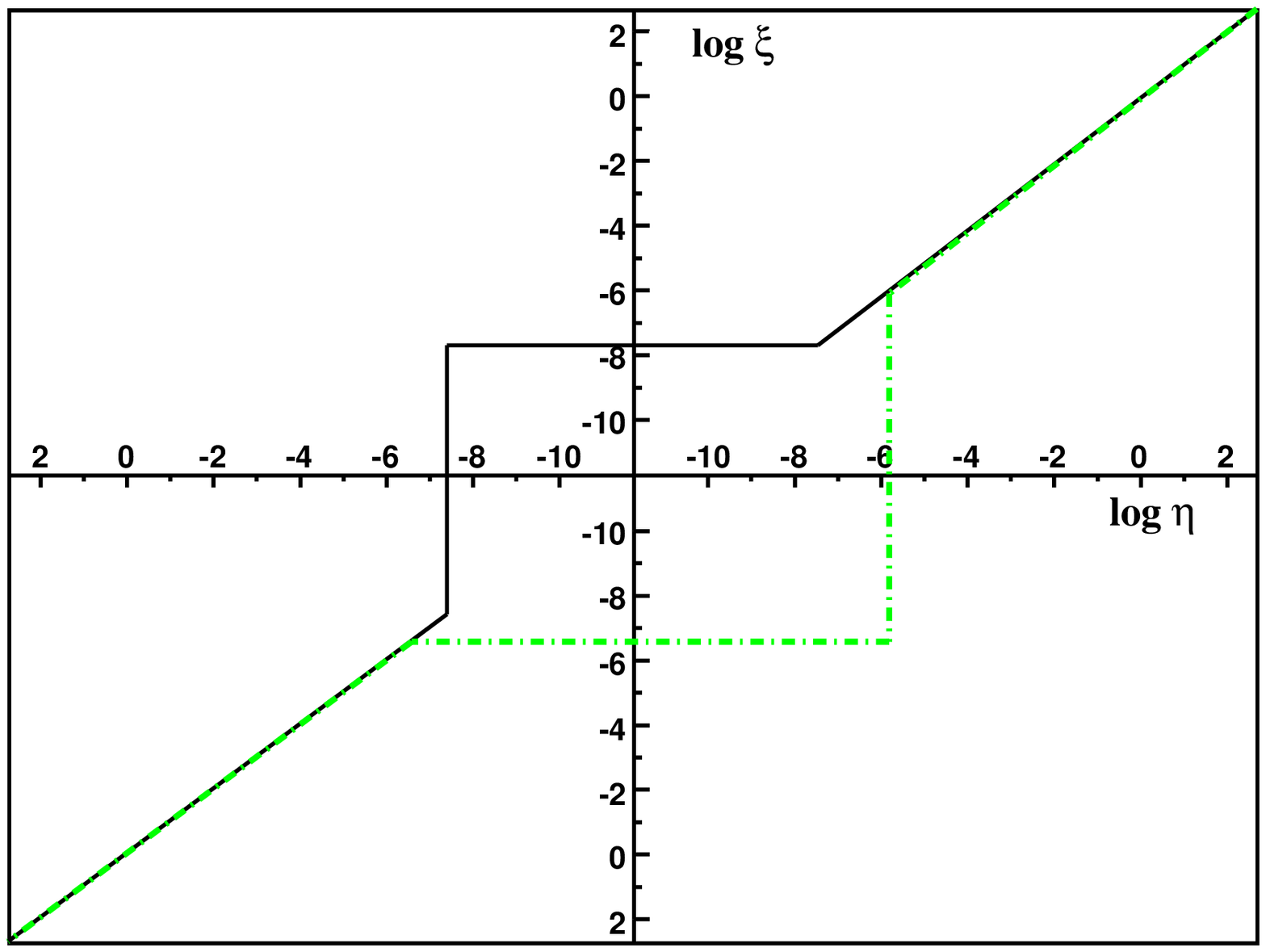}
 \caption{Left panel: $n=3$ LV. Right panel: $n=4$ LV. The LV parameter space is shown. The current best constraints (when they exist) are drawn in red. Black solid lines represent values of $(\eta, \xi)$ for which the $\gamma$-decay threshold $k_{\gamma-dec} \simeq 10^{19}~\eV$. Dot-dashed, green lines indicate pairs $(\eta,\xi)$ for which the pair production upper threshold $k_{\rm up} \simeq 10^{20}~\eV$.}
 \label{fig:constraints}
\end{figure}

Finally, we notice that these two methods do not in principle exclude each other (although the physical effects are mutually exclusive). This is the case if a lower limit to the photon flux at $10^{19}~\eV$ is imposed by the experiments (thus implying the absence of $\gamma$-decay), while the upper limit at $10^{20}$ eV is confirmed. The constraint obtained in this case would be very strong, as the allowed region is given by the intersection of the two regions bounded by the green dot-dashed and the black solid lines in Fig.~\ref{fig:constraints}. 
Figure~\ref{fig:constraints} shows that in this case, for $n=3$ one would get $|\xi^{(3)}|,|\eta^{(3)}|\lesssim 10^{-14}$, basically ruling out this model. Remarkably, also the case $n=4$ would be strongly constrained as in this case one could deduce $|\xi^{(4)}|,|\eta^{(4)}| \lesssim 10^{-6}$. 

This would be the first strong and robust limit on the $n=4$, CPT even LV QED, which, as we explained in the Introduction, section \ref{sec:intro}, is also favored from a theoretical point of view.
Accidentally, as a methodological remark, this result also shows that while much attention was focused on the detection of single events at GZK energies for constraining LV, full spectral information could be more effective. 

In conclusion, we think that these results fully show the crucial role that the PAO experiment will play in coming years as the main experiment for testing fundamental symmetries of nature to unprecedented precision levels.

\ack

We wish to thank A.~Celotti, D.~Mattingly and G.~Sigl for useful discussions and remarks.

\section*{References}


\end{document}